\pdfoutput=1
\documentclass[aps,prb,reprint,groupedaddress,longbibliography]{revtex4-1}
 
\usepackage{amssymb}
\usepackage{xcolor}
\usepackage{amsmath}
\usepackage{grffile}
\usepackage{cancel}
\usepackage{hyperref}

\hypersetup{
  pdfproducer= none,  
  pdfcreator = none,  
  colorlinks = true,
  allcolors = black, 
}

\def\beq{\begin{equation}}
\def\eeq{\end{equation}}
\def\bea{\begin{eqnarray}}
\def\eea{\end{eqnarray}}
\def\fb{f_{\rm b}}
\def\fbs{f_{\rm b}^{\rm s}}

\usepackage{amssymb}
\usepackage{color}
\usepackage{amsmath}
\usepackage{bm}
\usepackage[english]{babel}

\usepackage{graphicx}
\usepackage{tikz}
\usetikzlibrary{positioning}
\usepackage{ifthen}
\usepackage{xstring}

\newcommand{\Fig}[1]{Fig.~\ref{#1}}
\newcommand{\Figs}[1]{Figs.~\ref{#1}}
\newcommand{\Sec}[1]{Section~\ref{#1}}

\newcommand{\Tab}[1]{Table~\ref{#1}}

\newcommand{\Eqn}[1]{Eqn.~\ref{#1}}

\newcommand{\ebb}{\epsilon_\textrm{bb}}
\newcommand{\ebr}{\epsilon_\textrm{br}}
\newcommand{\err}{\epsilon_\textrm{rr}}

\newcommand{\coordination}{k}

\newcommand{\n}[1]{{\color{black} #1}}
\newcommand{\nn}[1]{{\color{black} #1}}

\definecolor{bookColor}{cmyk}{0 , 0  , 0   , 1.00}  
\color{bookColor}

\usepackage{array}
\newcolumntype{C}[1]{>{\centering\arraybackslash}p{#1}}

\color{bookColor}

\begin{document}


\title{Landscape of kinetically trapped binary assemblies}


\author{Ranjan V. Mannige}
\email[]{rvmannige@lbl.gov}
\affiliation{Molecular Foundry, Lawrence Berkeley National Laboratory, 1 Cyclotron Road, Berkeley, CA, U.S.A.}


\date{\today}

\begin{abstract}
\color{bookColor}
\n{For two-component assemblies, an inherent structure diagram (ISD) is the relationship between set inter-subunit energies and the types of kinetic traps (inherent structures) one may obtain from those energies. It has recently been shown that two-component ISDs are apportioned into regions or plateaux within which inherent structures display uniform features (e.g., stoichometries and morphologies). Interestingly, structures from one of the plateaux were also found to be robust outcomes of one type of non-equilibrium growth, which indicates the usefulness of the two-component ISD in predicting outcomes of some types of far-from-equilibrium growth. However, little is known as to how the ISD is apportioned into distinct plateaux. Also, while each plateau displays classes of structures that are morphologically distinct, little is known about the source of these distinct morphologies.}  
\n{This article outlines} an analytic treatment of the two-component ISD, 
and show\n{s} that the manner in which any ISD is apportioned arises from a single unitless order parameter. Additionally, the analytical framework allows for the characterization of local properties of the trapped structures within each \n{ISD} plateau. This work may prove to be useful in the design of novel classes of robust nonequilibrium assemblies.
\end{abstract}


\maketitle

\usetikzlibrary{positioning}

\newcommand{\simplex}[1][brbrb]{%
	\begin{tikzpicture}[thick, scale=.6, anchor=base, baseline=0.65ex, transform shape, mystyle/.style={clip, inner sep=0,outer sep=0}]
		\def \dark  {blue!100!white} 
		\def \light {red!40!white}
		\def \blank {white!20!white}
		%
		\edef\mya{1mm}
		\StrChar{#1}{2}[\compto] 
		\def \fill {\blank}
		\IfStrEq{\compto}{b}{\def \fill {\dark}}{} 
		\IfStrEq{\compto}{r}{\def \fill {\light}}{}
		\IfStrEq{\compto}{w}{\def \fill {\blank}}{}
		\IfStrEq{\compto}{-}{}{\filldraw[fill=\fill, draw=black, thick, inner sep=0, outer sep=0] (\mya*5.0,\mya*3.0) circle (\mya*1);} 
		\StrChar{#1}{3}[\compto]
		\IfStrEq{\compto}{b}{\def \fill {\dark}}{} 
		\IfStrEq{\compto}{r}{\def \fill {\light}}{}
		\IfStrEq{\compto}{w}{\def \fill {\blank}}{}
		\IfStrEq{\compto}{-}{}{\filldraw[fill=\fill, draw=black, thick, inner sep=0, outer sep=0] (\mya*3.0,\mya*5.0) circle (\mya*1);}
		\StrChar{#1}{4}[\compto]
		\IfStrEq{\compto}{b}{\def \fill {\dark}}{} 
		\IfStrEq{\compto}{r}{\def \fill {\light}}{}
		\IfStrEq{\compto}{w}{\def \fill {\blank}}{}
		\IfStrEq{\compto}{-}{}{\filldraw[fill=\fill, draw=black, thick, inner sep=0, outer sep=0] (\mya*1.0,\mya*3.0) circle (\mya*1);}
		\StrChar{#1}{5}[\compto]
		\IfStrEq{\compto}{b}{\def \fill {\dark}}{} 
		\IfStrEq{\compto}{r}{\def \fill {\light}}{}
		\IfStrEq{\compto}{w}{\def \fill {\blank}}{}
		\IfStrEq{\compto}{-}{}{\filldraw[fill=\fill, draw=black, thick, inner sep=0, outer sep=0] (\mya*3.0,\mya*1.0) circle (\mya*1);}
		\StrChar{#1}{1}[\compto]
		\IfStrEq{\compto}{b}{\def \fill {\dark}}{} 
		\IfStrEq{\compto}{r}{\def \fill {\light}}{}
		\IfStrEq{\compto}{w}{\def \fill {\blank}}{}
		\IfStrEq{\compto}{-}{}{\filldraw[fill=\fill, draw=black, thick, inner sep=0, outer sep=0] (\mya*3.0,\mya*3.0) circle (\mya*1);}
	\end{tikzpicture}%
}

\color{bookColor}

\section{Introduction}
An important goal of molecular \n{design is to facilitate} the \n{{\it robust}} self-assembly of \n{elaborate} structures from smaller components. Here, ``\n{robust}'' refers to the capacity of the same components starting off in varying \n{environmental} conditions to result in the same or similar outcome\cite{Bryngelson1995,Dill1997,Onuchic1997,Papoian2003,Wang2003,Nguyen2009}
The outcome of a \n{robust} molecular assembly process is traditionally thought to be the equilibrium \n{(lowest-free-energy)} structure, with non-equilibrium structures posing hurdles (kinetic traps) towards that process\n{\cite{Bryngelson1995,Dill1997,Faraldo-Gomez2007,Leopold1992,Oakley2011,Onuchic1997,Papoian2003,Wang2003,Nguyen2009,Markegard2015,Hedges2014}}. \n{Yet, some regimes of far-from-equilibrium growth have also been shown to result in robust outcomes\cite{Kim2009,Scarlett2010,Sue2015,Mannige2015b}, which indicates novel (non-equilibrium) routes towards building robust molecular assemblies.} 

\begin{figure}[t!]
\centering
\includegraphics[width=0.45\textwidth]{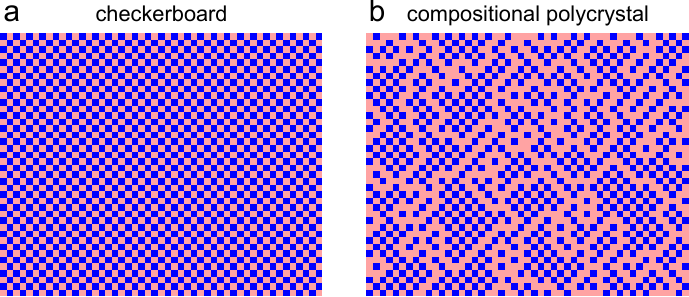} 
\caption{\textbf{Equilibrium versus nonequilibrium structures obtained from growth.} Given two components (``red'' and ``blue'') whose heterogeneous interaction energy is the lowest (i.e., $\ebr<\ebb,\err$), near-equilibrium growth results in a ``checkerboard'' patterned structure \n{with equal ratio of blues and reds ($f_\textrm{b}=0.5$), shown in (a) for a 2d square lattice} (henceforth, the `red' components are shown faded for increased contrast compared to blue components). 
Interestingly, at a particular \n{interaction energy regime} ($\ebr<\err\equiv0<2|\ebr|<\ebb$), 
while the equilibrium structure remains the checkerboard \n{(a)}, 
a new class of non-equilibrium structures -- \n{compositional} polycrystalline assemblies (b) \n{with a particular $f_\textrm{b}$ ($\sim0.364$)} -- 
form from a range of non-equilibrium growth regimes and red:blue ratios in solution\cite{Sue2015,Mannige2015b} \n{(also shown in \Fig{Fig:robustInherentSt}a,b)}. 
This particular class of polycrystalline structure is but one of a range of possible structures available within the inherent structure diagram (\Fig{Fig:robustInherentSt}\n{d}).\label{Fig:checkerVpoly}}
\end{figure}

\n{While non-equilibrium outcomes in relatively detailed and complex processes have been studied 
-- e.g., in the process of virus capsid assembly\cite{Nguyen2009} and DNA hybridization\cite{Markegard2015} -- 
computational limitations prevent more systematic studies of such systems. Alternatively, simpler ordered frameworks formed from two types of components have been useful in systematically distinguishing the outcomes of equilibrium and non-equilibrium regimes for nucleation and growth\cite{Sanz2007,Peters2009,Kim2009,Scarlett2010,Whitelam2012,Whitelam2014a,Sue2015,Mannige2015b}. 
This report considers growth post nucleation, where, depending on the rate of growth, both equilibrium and non-equilibrium results can be resolved\cite{Sue2015,Mannige2015b}. For convenience, this report refers to the two component types as ``red'' and ``blue'' (as is the convention in Refs.~\citenum{Sue2015} and~\citenum{Mannige2015b}).

Near-equilibrium growth is attained when the rate of growth is so slow that binding is nearly reversible. The first utility of two-component growth is that the outcomes of equilibrium growth are relatively easy to predict. For example, if the heterogeneous energy of interaction is dominantly the lowest in energy, analogous to the components sodium (Na$^+$) and chloride (Cl$^-$) in a crystal, the equilibrium outcome is ``binary'' (otherwise the outcomes dominantly display one or the other subunit). Here, binary indicates that each neighbor is surrounded by neighbors of opposite color. In a two-dimensional (2d) square lattice, such a binary structure resembles a checkerboard structure (\Fig{Fig:checkerVpoly}a). A convenient metric to characterize binary assemblies is the fraction of blues in the assembly ($f_\textrm{b}$), which, in the case of a binary solid is 0.5 (like in Na$^+$Cl$^-$). 

When growing an assembly at far-from-equilibrium rates (i.e., at much greater growth rates), the expected equilibrium outcomes (such as structures displaying $f_\textrm{b}=0.5$ for the system above) are not the outcomes that one encounters\cite{Sanz2007,Peters2009,Kim2009,Scarlett2010,Whitelam2014a,Sue2015,Mannige2015b}. 
This is because, at particular growth rates and subunit-subunit energetics, kinetic traps involving multiple components within the bulk of the assembly are not permitted to rearrange into equilibrium features before those defects get buried\cite{Whitelam2014}. 
Mostly, non-equilibrium outcomes are highly sensitive to solution conditions such as the combined and relative concentration of the two components in solution. However, in some cases, the non-equilibrium outcomes are actually robust to some growth conditions\cite{Kim2009,Scarlett2010,Sue2015,Mannige2015b}. 
Presently, while such robust yet non-equilibrium assemblies may present us with new ways to produce novel and controllable materials, an overarching framework for making predictions for such processes is lacking, especially because most existing frameworks for predicting molecular outcomes are based on equilibrium statistical mechanics (see discussions in Refs.~\citenum{Whitelam2012} and~\citenum{Whitelam2014a}).}

\begin{figure*}[t!]
\centering
\includegraphics[width=0.9\textwidth]{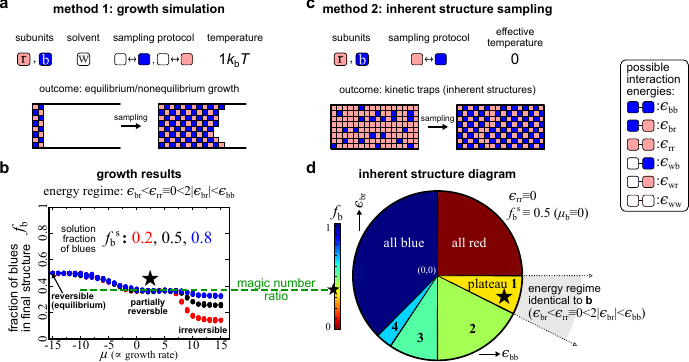} 
\caption{\n{\textbf{Previously discussed growth (a,b)\cite{Mannige2015b} and inherent structure sampling (c,d,)\cite{Sue2015}.} 
As replicated from a previous study\cite{Mannige2015b}, Monte Carlo simulations of growth (a; see \Sec{Method:growth}) at a certain energy regime result in assemblies whose composition (fractions of blues; $f_\textrm{b}$) remain robust to both a range of growth rates and relative solution concentrations ($\bigstar$ in b)\cite{Sue2015,Mannige2015b}. Inherent structure sampling of a random two-component lattice model (c; see \Sec{Method:inherentStructure}) results in uniform regions (plateaux) of interaction space within which inherent structures describe uniform $f_\textrm{b}$ (d)\cite{Sue2015}. Interestingly, the robust structures obtained from growth at partially reversible rates ($\bigstar$) are identical to those obtained from plateau {\bf 1} of the inherent structure diagram, regardless of a range of starting configurations\cite{Sue2015,Mannige2015b}. The structures represent a class of compositional polycrystals (\Fig{Fig:checkerVpoly}b).\label{Fig:robustInherentSt}}}
\end{figure*}



\n{Refs.~\citenum{Sue2015} and~\citenum{Mannige2015b}, which will be focused on in more detail here, describe a non-equilibrium growth scenario that yields assemblies that are robust to {\it multiple} solution properties: growth rate and solution stoichometries. The model assumes the following hierarchy of subunit-subunit interaction energies\cite{Sue2015,Mannige2015b}:} the red-blue interaction energy ($\ebr$) is the most favorable, the red-red interaction energy ($\err$) is negligible and intermediate in strength, and the blue-blue interaction energy ($\ebb$) is high enough to never permit its occurrence during growth/assembly \n{(all interaction energies in this manuscript are taken to be the equivalent of binding free energies in actual molecules)}. 
These stipulations equate to the hierarchy $\ebr<\err\equiv0<2|\ebr|<\ebb$. \n{Subunit growth was simulated on a lattice  (\Fig{Fig:robustInherentSt}a; also see \Sec{Method:growth}), which yielded in Refs.~\citenum{Sue2015} and~\citenum{Mannige2015b} the following results.}

\n{At slow, near-reversible, growth rates ($\mu$ is taken as a proxy), the expected equilibrium outcome is obtained ($\mu=-15$ always yields $f_\textrm{b}=0.5$;  \Fig{Fig:robustInherentSt}b). However, some faster growth rates (about $-2<\mu<8$; `$\bigstar$' in \Fig{Fig:robustInherentSt}b) yield $f_\textrm{b}$ values that deviate from the equilibrium $f_\textrm{b}$. Yet, these structures are still robust to {\it both} a range of growth rates and solution stoichiometries\cite{Sue2015,Mannige2015b}. Robustness to growth rate ($\propto \mu$) is evident in the presence of a plateau ($\bigstar$) per curve in \Fig{Fig:robustInherentSt}b; Robustness to solution stoichiometry is evident in the adherence of all `$\bigstar$' plateaux to single magic number ($f_\textrm{b}\approx0.364$), regardless of the distinct ratios of red:blue set in in solution ($f_\textrm{b}^\textrm{s}=0.2,0.5,0.8$). The plateau `$\bigstar$' describes a range of growth rates that are neither reversible nor irreversible to the chosen interactions energies: they describe a partially reversible regime of growth where some types of binding are allowed to reverse (e.g., r-r) while some bind irreversibly (e.g., r-b) \cite{Sue2015,Mannige2015b}. The nonequilibrium assemblies may be termed as ``compositional polycrystals''\cite{Sue2015,Mannige2015b}} whose ``grains'' are patches of equilibrium regions (checkerboard) and whose boundaries constitute only red components (\Fig{Fig:checkerVpoly}b). \n{The word ``compositional'' in this term is meant to indicate that, ignoring color, the framework would be fully occupied and ordered\cite{Sue2015,Mannige2015b}.}

Interestingly, these polycrystalline structures obtained via partially reversible growth \n{are also obtained {\it via} a simpler sampling protocol (\Fig{Fig:robustInherentSt}c; see \Sec{Method:inherentStructure})\cite{Sue2015,Mannige2015b} that samples for inherent structures\cite{Stillinger1983,Stillinger1988,Bertin2005}. Inherent structures were obtained by a two-state sampling technique that allows only energetically favorable or neutral color switches (\Fig{Fig:robustInherentSt}c). Inherent structure sampling was applied to a range of random initial configurations at particular points in ($\ebb$,$\ebr$)-space ($\err\equiv0$ with no loss of generality\cite{Sue2015}). The resulting average $f_\textrm{b}$ as a function of ($\ebb$,$\ebr$), which may be referred to as an inherent structure diagram (ISD), is shown in \Fig{Fig:robustInherentSt}d. Interestingly, the ISD is apportioned into discrete regions (plateaux) within which the structures all describe nearly identical $f_\textrm{b}$. The region within the ISD that describes plateau 1 describes structures that are identical to those found from partially reversible growth ($\bigstar$ in \Fig{Fig:robustInherentSt}b): structures within plateau 1 are compositional polycrystals (\Fig{Fig:checkerVpoly}b) whose $f_\textrm{b}$s are identical to those obtained from partially reversible growth. Additionally, all points within plateau 1 exclusively satisfy the interaction energy relationship that was used to grow the non-equilibrium structures (\Fig{Fig:robustInherentSt}b), namely, $\ebr<\err\equiv0<2|\ebr|<\ebb$. This indicates a substantial connection between the structures available within an inherent structure diagram and the outcomes of some types of non-equilibrium growth; the other plateaux in the inherent structure diagram -- each describing potential materials with specific local structural properties (\Fig{Fig:plateaux}) -- may, too, lead to the nonequilibrium growth of other materials with other magic numbers.}

\begin{figure}[t!]
\centering
\includegraphics[width=0.44\textwidth]{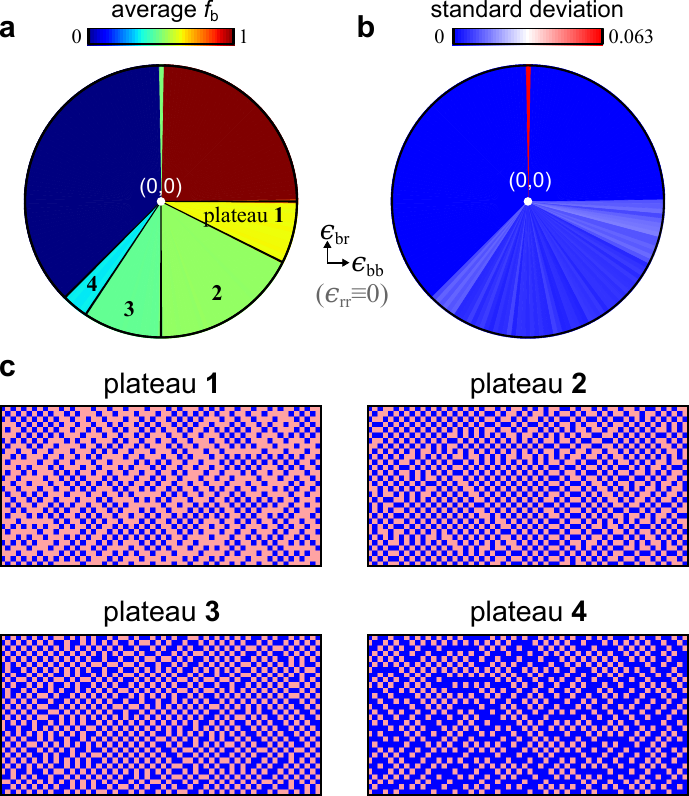} 
\caption{\n{\textbf{Each plateau in (a) displays homogeneous $f_\textrm{b}$ (b) and distinct classes of polycrystalline inherent structure (c).} 
It has already been shown that inherent structures observed in plateau 1 of the inherent structure diagram or ISD (a; \Fig{Fig:robustInherentSt}d) represent compositional polycrystals with binary domains/grains and all-red grain boundaries\cite{Sue2015}. Interestingly, all other non-trivial regions of the ISD (plateaux 2 through 4) also describe distinct classes of compositional polycrystals, whose structures are distinguished by distinct grain boundaries (e.g., plateau 4 describes boundaries that are all blue). As indicated by (b), while each plateau describes a range of compositional polycrystals, each structure within the same plateau describes the same $f_\textrm{b}$. \label{Fig:plateaux}}}
\end{figure}

While the potential exists for inherent structures to explain the outcomes of \n{two}-component out-of-equilibrium assemblies, little is known about what determines the plateaux evident in the ISD. \n{This manuscript describes a} pen-and-paper model \n{that will be used} to show that each ISD plateau, as well as the properties (local environments) of the inherent structures within each plateau, are consequences of the local connectivity (degree) of the assembly framework (lattice). 
Specifically, the formalism developed below predicts {\it i)} the boundaries of potential plateaux in any two-component ISD, and {\it ii)} local environments of each plateau-specific inherent structure. This report also shows that the local rules developed here can not predict the actual magic ratios ($f_\textrm{b}$) each plateau displays, as those are a matter of global (not local) lattice connectivity\cite{Sue2015}.


\section{\label{PreviousMethods}Numerical Methods Described Previously\cite{Sue2015,Mannige2015b}}

\n{This section is meant to accompany the discussions in the introduction pertaining to the correspondence between outcomes of non-equilibrium growth and inherent structure sampling. }

\subsection{Interaction graphs used in the Monte Carlo sampling}
Any arbitrary interaction network (graph) may be used in our sampling procedure below. Particular graphs used here are the periodically connected square lattices in dimension $d$ ($d\in\{1,2,\dotsc,5\}$) and the \textbf{nbo} lattice. 
The \textbf{nbo} lattice\cite{Norman1997} represents the \n{interaction network observed in a metal-organic framework (MOF-2000) described before\cite{Sue2015}. \nn{The periodic cell of the \textbf{nbo} graph is as follows: \newline
\textcolor{white}{.}\hfil\includegraphics[width=0.1\textwidth]{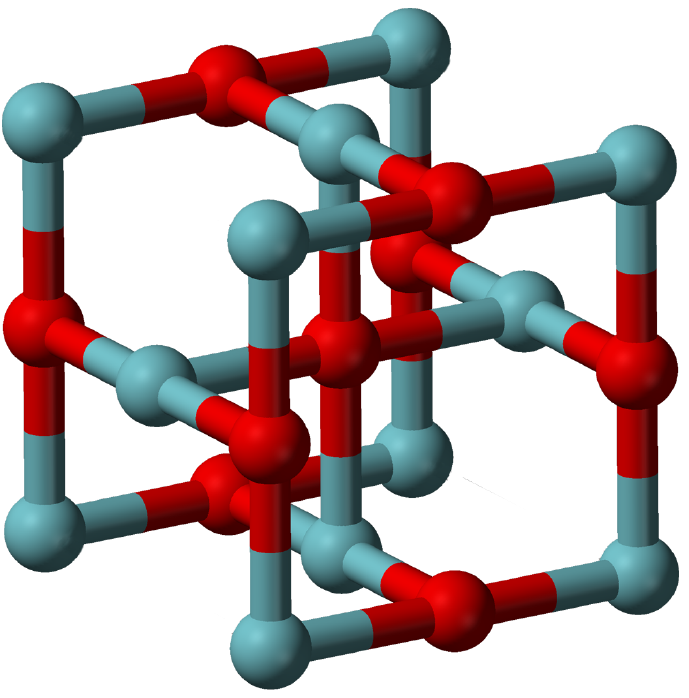} \hfil\newline
\textcolor{white}{.}\hfil({\bf{nbo}} topology; image source: Wikipedia\footnote{Wikipedia URL:\newline \href{https://en.wikipedia.org/wiki/Niobium_monoxide}{https://en.wikipedia.org/wiki/Niobium\_monoxide}})} \hfil\newline}
Note that the higher dimension square lattices ($d>3$), although unnatural, are useful to understand materials characterized by higher degrees or coordination number.

\subsection{\n{Method 1: Monte Carlo sampling algorithm for nucleated growth (\Fig{Fig:robustInherentSt}a)\cite{Mannige2015b}.}\label{Method:growth}} 
\n{The sluggish dynamics of particles within a solid is a primary cause of kinetic trapping within a growing multicomponent assembly. Due to this simple physical origin, kinetic trapping from growth can be reproduced by simple physical models that account for this slow dynamic\cite{Sanz2007,Peters2009,Kim2009,Scarlett2010}. This section discusses a lattice model of growth similar to the models used in Refs.\citenum{Whitelam2014a,Hedges2014,Sue2015} and identical to that used in Ref.~\citenum{Mannige2015b}. As sketched in \Fig{Fig:robustInherentSt}a, each site on the lattice may be either occupied by a subunit/component colored red or blue or by solvent that is colored white. The energy of the system is
\begin{equation}
\displaystyle
E = \sum_{i,j}^\textrm{edges} \epsilon_{C(i)C(j)} + \sum_{i}^\textrm{nodes} \mu_{C(i)}. \label{energy}
\end{equation}

The first sum iterates through all nearest-neighbor interactions. Here $C(i) = \textrm{b}$, $\textrm{r}$ or $\textrm{w}$ if node $i$ is `blue', `red' or `white', respectively. $\epsilon_{C(i)C(j)}$ is the pairwise free energy of interaction between two adjacent nodes $i$ and $j$ (simply called interaction `interaction energy' henceforth); therefore, there are six types of interaction energies: $\epsilon_{\textrm{b}\textrm{b}}$, $\epsilon_{\textrm{b}\textrm{r}}$, $\epsilon_{\textrm{r}\textrm{r}}$, $\epsilon_{\textrm{w}\textrm{b}}$, $\epsilon_{\textrm{w}\textrm{r}}$, and $\epsilon_{\textrm{w}\textrm{w}}$.

The second iterates through all nodes, which represent subunits (b,r) and solvent (w). The chemical potential term $\mu_{C(i)}$ is $\mu$, $-\ln(f_\textrm{b}^\textrm{s})$ and $-\ln(1-f_\textrm{b}^\textrm{s})$ for w, b and r, respectively.

In the absence of pairwise energetic interactions (i.e. in implicit solution), the likelihood that a given site will be white, blue or red is respectively $\left\{p_{\rm w}, p_{\rm b}, p_{\rm r} \right\} = \left\{{\rm e}^{-\mu},\fb^\textrm{s},1-\fb^\textrm{s}\right\} \left(1+{\rm e}^{-\mu} \right)^{-1}$.

The Monte Carlo growth simulations were performed in the following manner. The simulation box was 400 sites wide by 40 sites high. The first six columns were populated with the equilibrium checkerboard structure. Our sampling protocol involved selecting a node at random and attempting to change the color of that node. Given a white node, an attempt is made to change its color to either red or blue, and given a red/blue colored site, an attempt is made to change its color to white. 

When attempting to switch from white to colored, blue is chosen with probability $\fb^\textrm{s}$, and red is chosen otherwise. Red-to-blue switches are not allowed, which mimics the idea that binding and unbinding events are the dominant way in which the configurational degrees of freedom evolve.

To maintain detailed balance with respect to the \Eqn{energy}, the acceptance rates for these moves were as follows ($\Delta E$ is the energy change resulting from the proposed move):
\begin{eqnarray}
\textrm{r} \to \textrm{w} &:& \min(1,(1 - \fb^\textrm{s})\exp[-\Delta E]); \nonumber \\
\textrm{w} \to \textrm{r} &:& \min(1,(1 - \fb^\textrm{s})^{-1}\exp[-\Delta E]); \nonumber \\
\textrm{b} \to \textrm{w} &:& \min(1,\fb^\textrm{s} \exp[-\Delta E] ); \nonumber \\
\textrm{w} \to \textrm{b} &:& \min(1,(\fb^\textrm{s})^{-1} \exp[-\Delta E]  ). \nonumber
\end{eqnarray}
The implicit solution abundances of red and blue are controlled by the chemical potential term that appears in \Eqn{energy} and therefore in the term $\Delta E$. Our choice to insert blue particles with likelihood $\fbs$ does not by itself result in a thermodynamic bias for one color over the other (because this bias in proposal rate is countered by the non-exponential factors in the acceptance rates). Instead, insertions are biased so that the dynamics of association are consistent with the thermodynamics of the model. For instance, if blue particles are more numerous in solution than red ones, it is physically appropriate to insert blue particles into the simulation box more frequently than red particles. Consider the limit of large positive $\mu$: the `solid solution' that results as the box fills irreversibly with colored particles will have a red:blue stoichiometry equal to that of the notional solution only if blue particles are inserted with likelihood $\fbs$. As a technical note, the chemical potential term present in $\Delta E$ ends up simply canceling the non-exponential factors in the acceptance rates, but Ref.~\citenum{Mannige2015b} chose to write acceptance rates as shown in order to make clear which pieces are imposed by thermodynamics, and which pieces are chosen for dynamical reasons. Note that temperature is not defined explicitly, but can be considered to be subsumed into the energetic parameters of the model.

The parameter values used to obtain \Fig{Fig:robustInherentSt}d in $k_\textrm{b}T$ are $\epsilon_\textrm{bb}=70$,$\epsilon_\textrm{br}=-7$,$\epsilon_\textrm{rr}=0$, which is replicated from Ref.~\citenum{Mannige2015b}.}

\subsection{\n{Method 2:} Sampling algorithm to obtain inherent structures (\Fig{Fig:robustInherentSt}c)\cite{Sue2015}.\label{Method:inherentStructure}} 
\n{As in Method 1, our energy function follows that of \Eqn{energy}, with the exception that white (solvent) sites are not allowed (\Fig{Fig:robustInherentSt}c).} Therefore, there are three types of interaction \n{(free)} energies: $\epsilon_{\textrm{b}\textrm{b}}$, $\epsilon_{\textrm{b}\textrm{r}}$, $\epsilon_{\textrm{r}\textrm{r}}$. The chemical potential term $\mu_{C(i)}$ is $0$ and $\mu_{\textrm{b}}$ for red and blue nodes, respectively. \n{These chemical potentials indicate how easy it is to switch to red and blue states based on the relative concentration of reds vs blues in a notional or implicit ``solution'' (the grand canonical reservoir of colored blocks).} Therefore, the fraction of blue nodes in implicit solution is
\begin{equation}
\displaystyle
f_\textrm{b}^\textrm{s} = \frac{e^{\mu_\textrm{b}}}{(1+e^{\mu_\textrm{b}})}.\label{eq:fbsol}
\end{equation}

\n{Inherent structure sampling  happens in the following fashion. All sites within the lattice are randomly populated with subunits (either red or blue). At each sampling step, state changes are attempted by randomly selecting a site and then attempting a switch in the subunit's color. If the switch is energetically favorable or neutral, then that state change is kept. This process continues until no state change occurs. While these rules in effect represent sampling at $T=0$, they are also equivalent to a sampling protocol where temperature is finite (say $T=1$) and where the interaction energies are multiplied by $\infty$.

All of the inherent structure diagrams or ISDs discussed in this paper (\Fig{Fig:robustInherentSt}c, \Fig{Fig:plateaux}a-b, \Fig{Fig:plateaux}a-c) are calculated with $\mu_\textrm{b}\equiv0$, i.e., relative concentrations of red versus blue subunits in the notional solution are equal ($f_\textrm{b}^\textrm{s}\equiv0.5$). Changing the relative concentration (by tuning $\mu_\textrm{b}$) effectively shifts or translates the features (plateaux) within the ISD with respect to $\ebb$ and $\err$ without changing the positions of the plateaux relative to each other (see Fig.~3B in Ref.~\citenum{Sue2015}).

As calculated previously\cite{Sue2015,Mannige2015b}, each point on the inherent structure diagram obtained in \Fig{Fig:robustInherentSt}b represents the average fraction of blues ($f_\textrm{b}$) of a number of inherent structures, calculated using the method shown above, at that particular ($\epsilon_\textrm{bb}$,$\epsilon_\textrm{br}$) point.

A theoretical framework is explored below that attempts to chart such inherent structure diagrams (ISDs).
}

\section{Theory: Charting the Inherent Structure Diagram}

\textbf{Possible ISD plateau boundaries from simplex transitions.} First, \n{this report} shows that borders between plateaux may be surmised by understanding the behavior of color switches within simpler sub-structures of the lattice. Such units of structure are described as ``simplices'', which comprise a central node and all of its nearest neighbors. \n{While the term ``simplex'' is traditionally considered to be extensions of a triangle in a given dimension, this report utilizes the algebraic topology definition of a simplex (abstract simplicial complex), in which the word ``simplex'' means a finite (but still useful and basic) set of vertices. For the purpose of this paper, a simplex is a convenient word to be used repeatedly instead of the longer yet appropriate term ``central node with all its neighboring nodes''.} 

While simplices can be described for any regular graph (including random regular graphs), the derivation below will be introduced using the square lattice/graph. For square lattices in dimension $d$, each simplex will have $2d+1$ nodes and is denoted graphically as a star graph (\simplex[brrrr]), with the different color combinations (e.g., \simplex[brbrr], \simplex[rrbrb], ...) describing the various degrees of freedom for the simplex. In these simplices, the central node is the only node on which a color-switch will be attempted. For any specific set of peripheral nodes (say \simplex[-rbrb]), ``state \n{b}'' and ``state \n{r}'' are defined as simplices whose central node is of color ``b'' and ``r'', respectively. So, for the peripheral configuration ``\simplex[-rbrb]'', ``state \n{r}'' is ``\simplex[rrbrb]'' and ``state \n{b}'' is ``\simplex[brbrb]''. There are $2d+1$ number of simplex pairs that describe energetically unique environments, each corresponding to the a unique number $n_\textrm{b}$ of ``b''s that surround a central node ($n_\textrm{b} \in \{0,1\dotsc,2d\}$). Note that, energetically, the exact configuration of neighbors do not matter since their energies are identical; so, \simplex[brbrb] and \simplex[bbbrr] are equivalent simplices. Important to our treatment, each simplex pair can be treated as a single-site color switch conceptually occurring somewhere in the infinite assembly.

{\bf Establishing necessary criteria for ISD plateau boundaries.} Given a sampling algorithm, inherent structures are ``stuck'' conformations, i.e., zero degrees of freedom are available to any site in such structures (unless temperature is added to the system). Therefore, any plateau boundary in the ISD must lie at regions of ($\ebb,\ebr$)-space where at least one simplex freely switches back and forth (e.g., \simplex[brbrb]$\leftrightarrow$\simplex[rrbrb]), as such a boundary would indicate that adjacent regions in the energy space cause locally irreversible switches for that simplex (\simplex[brbrb]$\leftarrow$\simplex[rrbrb] or \simplex[brbrb]$\rightarrow$\simplex[rrbrb]), which is a precursor to a jammed or ``stuck'' conformation. Not all such moves are likely to cause jammed conformations, as peripheral sites may also be allowed to evolve to ``unjam'' the simplex in question. However, such equalities are necessary criteria for ISD boundaries, whose closed forms will be derived below.

\begin{figure}[t!]
\centering
\includegraphics[width=0.35\textwidth]{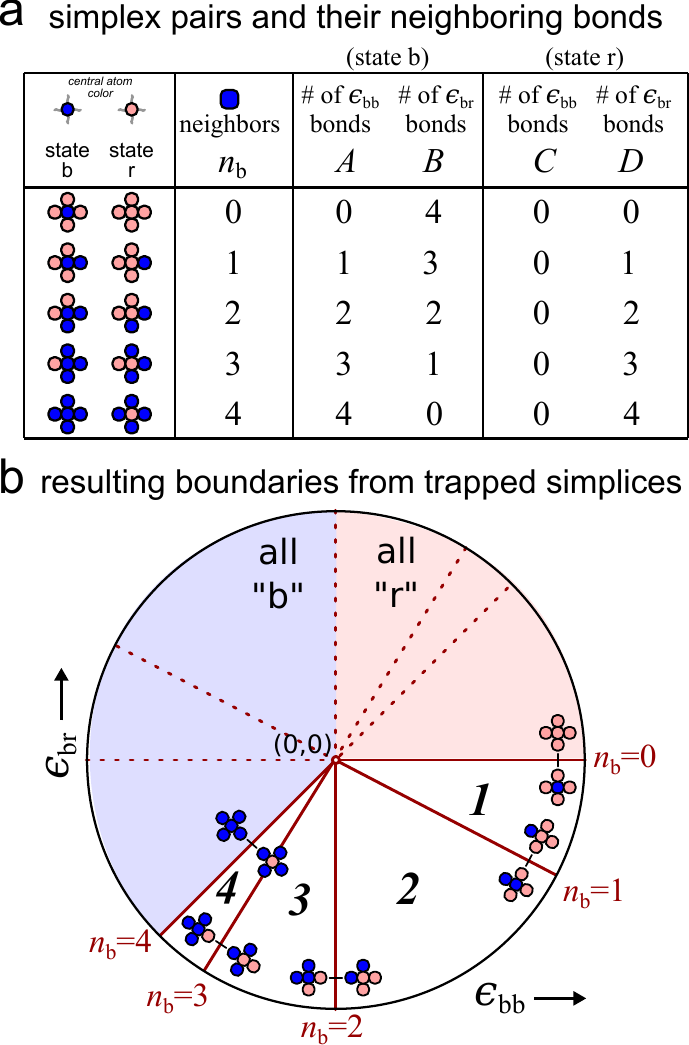} 
\caption{\textbf{Obtaining a closed-form description of plateau boundaries in \textit{d}=2.} Plateau boundaries may be deciphered by comparing for each simplex pair (state r and b) the number of b-b and b-r bonds available to each state (a). Applying these numbers to the basic transition equation (\Eqn{Eq:border1}) results in borders that depend only on the number of surrounding `b's ($n_\textrm{b}$), solution concentrations of each component ($\propto \mu_\textrm{b}$), and the connectivity ($\coordination=2d$) of the lattice (\Eqn{Eq:border2}). Plotting \Eqn{Eq:border2} results in plateaux (b) that are identical to those obtained from inherent structure sampling (\Fig{Fig:robustInherentSt}a). Correspondence in other lattices is also observed in \Fig{Fig:plateaux}.\label{Fig:boundaries}}
\end{figure}

\begin{figure}[t!]
\centering
\includegraphics[width=0.5\textwidth]{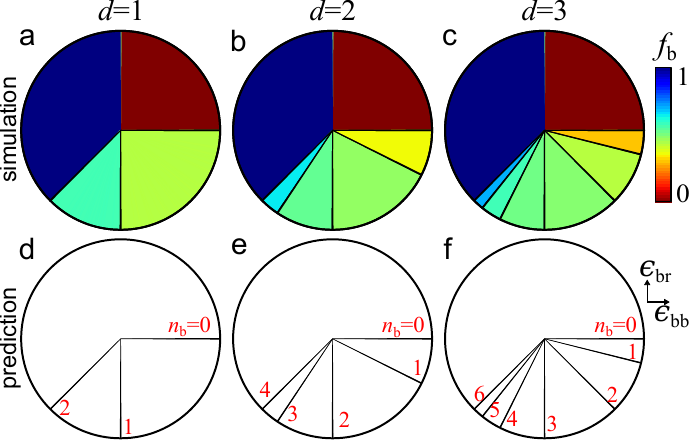} 
\caption{{\bf{Plateau boundaries obtained from simulation (a-c) and \Eqn{Eq:border2} (d-f) coincide within the antiferromagnetic regime.}} Simulations were performed for square lattices of varying dimension ($d$). For any ($\ebb,\ebr$)-pair, the starting conditions were chosen to be a random mixture of equal parts blue and red and $f_\textrm{b}^\textrm{sol}=0.5$ ($\mu_\textrm{b}=0$); the relevant plateaux (with $0<f_\textrm{b}<1$) remain stable regardless of the starting conditions\cite{Sue2015}. The antiferromagnetic regions in an ISD (a-c) display inherent structures that are neither all blue nor all red.
\label{Fig:plateaux}}
\end{figure}

\begin{figure*}
\centering
\includegraphics[width=0.9\textwidth]{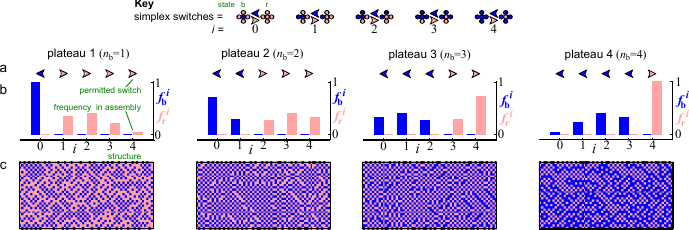} 
\caption{{\bf{Local structure evident within a plateau \n{(b,c)} are direct outcomes of energetically downhill simplex switching rules \n{(a)}.}} \n{Simplex switching involves swapping the color of the central node of a simplex. The key indicates possible simplex switches, where the simplex pair is defined by a specific number of blues in the neighborhood ($i$; see Definitions). Panel (a) describes the allowed moves within simplices of particular neighborhood $i$, where the blue and light red arrows respectively indicate a simplex switch towards state `b' and `r'. Given that only energetically downhill switches are allowed, these unidirectional switching rules are hard constraints on the types of simplices that are allowed within an inherent structure, which is measured in (b) by $f_\textrm{b}^i$ and $f_\textrm{r}^i$. ($f_\textrm{x}^i$ is the number of `x' sites that are surrounded by $i$ number of blue sites divided by the total number of `x' sites.) $f_\textrm{r}^{i} = 0$ if $i < n_\textrm{b}$ and $f_\textrm{b}^{i} = 0$ if $i \geq n_\textrm{b}$ where $n_\textrm{b}$ is the plateau number.}  
This, in turn, dictates the specific types of compositionally polycrystalline materials that arise from each plateau (\n{c}), which are all marked by grains composed of alternating red-blue coloring, but whose grain boundaries are plateau-specific (going from all-red for plateau 1 to all-blue plateau 4, and distinct mixtures in between). Local environments for other dimensions and lattices are shown in \Fig{Fig:localStructureLarge}.\label{Fig:structures}}
\end{figure*}

For any simplex pair, one can assess whether a given energy regime would allow for free switching (\simplex[brbrb]$\leftrightarrow$\simplex[rrbrb]), or switching only in one direction (\simplex[brbrb]$\leftarrow$\simplex[rrbrb] or \simplex[brbrb]$\rightarrow$\simplex[rrbrb]), by comparing the difference between the interaction energy of the state \n{b} simplex versus the state \n{r} simplex. In particular, the following relationship, if met, allows for free switching, and if not met results in switching permitted only in one direction:
\begin{align}
\textrm{energy(state \n{b})} & = \textrm{energy(state \n{r})} \\
\textrm{e.g., energy(\simplex[brbrb])} & = \textrm{energy(\simplex[rrbrb])} \nonumber
\end{align}
For any pair with $n_\textrm{b}$ blue neighbors, and $\err\equiv0$, this relationship becomes:
\begin{equation}
A \ebb + B \ebr + \mu_\textrm{b} = C \ebb + D \ebr\label{Eq:border1}
\end{equation}
where $\mu_\textrm{b}$ is the chemical potential for a node with color `b', which dictates the value of the relative fraction of `b' in solution or $f_\textrm{b}^\textrm{sol}$ (see \Sec{Method:inherentStructure}); $A$ and $C$ are the number of  $\ebb$ bonds in simplex states r and b, respectively, and $B$ and $D$ are the number of $\ebr$ bonds in states r and b, respectively. A simplex pair with $n_\textrm{b}$ number of ``b'' neighbors yields
\begin{align}
A & = n_\textrm{b}; & B & = \coordination-n_\textrm{b}; & C & =0; & D & = n_\textrm{b} \label{eq:hijk} 
\end{align}
See, e.g., \Fig{Fig:boundaries}a for an example of such assignments. $n_\textrm{b}$ is a boundary-specific value set by the number of peripheral `b's ($n_\textrm{b}$) displayed by the specific simplex pair. The additional parameter $\coordination$ is the degree of the regular graph (and so $\coordination=2d$ for any square lattice in dimension $d$). Applying Eqns.~\ref{eq:hijk} to \Eqn{Eq:border1} gives
\begin{align}
\ebr = \frac{n_\textrm{b}}{(2n_\textrm{b} - \coordination)}\ebb + \frac{\mu_\textrm{b}}{(2n_\textrm{b} - \coordination)} \label{Eq:border2}
\end{align} 
This is the equation for all possible plateau borders within the antiferromagnetic regime of the ISD (defined by those regions in \Fig{Fig:boundaries}b whose structures are not all-red and all-blue; i.e., demarcated by simultaneous conditions $\ebr < \err\equiv 0$ and  $\ebr < \ebb$). There can only be $\coordination+1$ number of such boundaries, as $n_\textrm{b}$ can be only one of $\coordination+1$ values ($n_\textrm{b} = \{0,1,\dotsc,\coordination\}$). Aside from system properties ($\coordination$,$\mu_\textrm{b}$), each boundary is defined by the integer $n_\textrm{b}$ (\Fig{Fig:boundaries}b). Additionally, a possible plateau is defined by adjacent $n_\textrm{b}$s, with the adjacent boundaries being $n_\textrm{b}-1$ and $n_\textrm{b}$ for $n_\textrm{b}\leq k$ (\Fig{Fig:boundaries}b). 
Attesting to the utility of the ``simplex method'' above, the plateau boundaries obtained from our sampling methodology are precisely marked by boundaries indicated by \Eqn{Eq:border2}. Compare, e.g., plateaux obtained from simulation (\Fig{Fig:robustInherentSt}a) and \Eqn{Eq:border2} in (\Fig{Fig:boundaries}b); similar comparisons for other dimensions are shown in \Fig{Fig:plateaux}.

\textbf{Local structures associated with each plateau.} 
\n{This section shows} that the simplex formalism discussed above may be used to predict the types of inherent structures displayed by each plateau. 
\Fig{Fig:structures} displays structural properties and a structure for each of the plateaux in $d=2$ (\Fig{Fig:boundaries}b). \n{For each column that describes a distinct plateau, Panel (a)} describes the allowed (downhill) moves available to our sampling algorithm (at $T=0$)\n{. Panel (b)} describes the local environments displayed by the structure, as measured by \n{$f_\textrm{b}^i$ and $f_\textrm{r}^i$ (these are are the respective probabilities that a blue and red node is surrounded by $i$ number of blue neighbors; $i\in\{0,\dotsc,k\}$). The complete collection of $f_\textrm{b}^i$ and $f_\textrm{r}^i$ completely describes how the structure is locally distributed.} Finally, an example of the resulting polycrystalline structure is shown in \Fig{Fig:structures}c, and the relationships between interaction energies within each plateau is shown in \Tab{growthswitchesMS}.

\begin{figure*}
\includegraphics[width=0.8\textwidth]{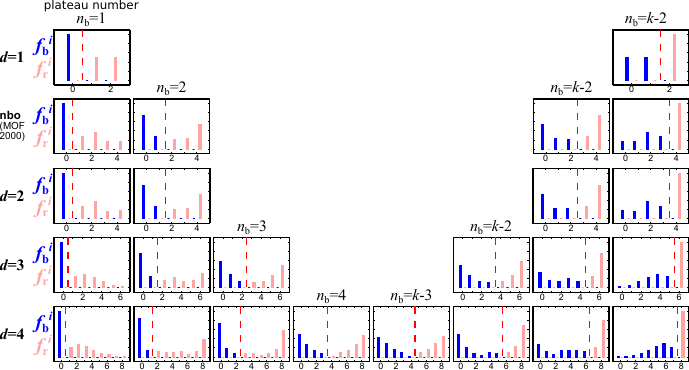} 
\caption{\n{As shown in \Fig{Fig:structures}b, panels within each row describe the local environment within each ISD plateaus obtained from a particular lattice. The lattices shown are square lattices of dimension $d=\{1,2,...,5\}$ ($k=2d$) and the \textbf{nbo} lattice ($k=4$). These graphs display two overarching constraints that are plateau-specific: $f_\textrm{b}^{i}=0 \textrm{~~if~~} i \geq n_\textrm{b}$ (see \Eqn{Eq:LocalRule1}) and $f_\textrm{r}^{i}=0 \textrm{~~if~~} i < n_\textrm{b}$  (see \Eqn{Eq:LocalRule2}). These hard limits are due to the plateau number ($n_\textrm{b}$) and can be visualized within each graph (b) as a vertical barrier (red dashed line) that separates the blue bars (to the left) from the light red bars (to the right).}
\label{Fig:localStructureLarge}}
\end{figure*}

\textbf{Polycrystalline types dictated by $\bm{n_\textrm{b}}$.}
Plateaux within the antiferromagnetic regime are associated with not one structure but an ensemble of configurations that display specific distributions of local environments. 
Interestingly, as suggested by the plateau-specific simplex switching rules in \Fig{Fig:structures}a, crucial aspects of the distribution of \n{$f_\textrm{b}^i$} and \n{$f_\textrm{r}^i$} depend primarily on the characteristic $n_\textrm{b}$ that describes the plateau:
\begin{align}
\n{f_\textrm{b}^i}=0 & \textrm{~~if~~} \n{i} \geq n_\textrm{b} \label{Eq:LocalRule1}\\
\n{f_\textrm{r}^i}=0 & \textrm{~~if~~} \n{i}  <   n_\textrm{b} \label{Eq:LocalRule2}
\end{align}
These vanishing probabilities are critical rules for understanding the structural features observed in each plateau; for example, \n{$f_\textrm{b}^{i\geq1}=0$} and \n{$f_\textrm{r}^{i{<}1}=0$} for plateau 1 ($n_\textrm{b}=1$) ensure the observed polycrystalline structure (\Fig{Fig:structures}, first column)\cite{Sue2015,Mannige2015b}, where red-blue binary crystalline domains are separated by purely red boundaries. The simplex method appears to therefore be useful to both apportion  the ISD into plateaux (\Figs{Fig:boundaries},~\ref{Fig:plateaux}), {\it and} account for the qualitative features of the material emerging from such plateaux (\Fig{Fig:structures}). Given that these predictions are based on local connectivity only, our boundary predictions also hold for higher dimensions (\Fig{Fig:plateaux}).

It is important to note that all of our results are dependent on simplices, which are local snapshots of the larger assembly. While this treatment is useful to motivate the existence of ISD plateaux and establish \n{constraints on the local neighborhoods} within plateau-specific inherent structures, it appears as though processes at the global network level dictate the {\it exact} magic ratios observed in each plateau. This is most evident when comparing two networks of identical degree or coordination ($\coordination$): the \textbf{nbo} lattice and $d=2$ square lattice \n{(rows two and three in \Fig{Fig:localStructureLarge})}, both of which have $\coordination=4$. While our method correctly predicts the plateau boundaries and constrains on local structure that result in the specific class of compositional polycrystal, the magic-number ratios are different in both networks (they are respectively $\sim 0.\bar{3}$ and $\sim 0.364$ for the \textbf{nbo} and 2d square lattice)\cite{Sue2015,Mannige2015b}. This arises from the fact that specific distributions within the unconstrained regions of the neighborhood vary between lattice types (\n{rows two and three in \Fig{Fig:localStructureLarge}}). This indicates that while the local connectivity describes the {\it type} of (polycrystalline) structure within each plateau, the actual ratio of reds:blues in solution are \n{the} result of non-local events as well; the exact effect of global connectivity on the magic \n{number} ratios \n{is} left for a future report.

\begin{table}[t] 
\begin{tabular}{| c | c | c | c |}
\hline 
Plateau & \multicolumn{3}{c|}{Interaction energy constraints} \\
\cline{2-4}
\#& 1 & 2 & 3 \\
\hline 
1   
&  $\err=0$ 
&  $\ebr < 0\ebb$ 
&  $-\frac{1}{2}\ebb < \ebr$ 
\\ 
2
&  $\err=0$ 
&  $\ebr < -\frac{1}{2}\ebb$
&  $0\ebr < \ebb$ 
\\
3 
& $\err=0$ 
& $\ebb < 0\ebr$ 
& $\ebr < \frac{3}{2}\ebb$ 
\\
4   
& $\err=0$ 
& $\frac{3}{2}\ebb < \ebr$ 
& $\ebr < \ebb$
\\
\hline
\end{tabular}
\caption{\textbf{Allowed interaction energy relationships within each plateau for \textit{d}=2.} Relationships obtained from \Eqn{Eq:border2} that place constraints on the inter-subunit interaction energies that account for the plateaux in \Fig{Fig:boundaries}.
\label{growthswitchesMS}}
\end{table}

\textbf{Allowing solution compositions to vary.} The last term containing $\mu_\textrm{b}$ in \Eqn{Eq:border2} indicates that changing the relative implicit solution concentration of components from 1:1 (i.e., setting $\mu_\textrm{b}\neq0$) will result in simply the skewing of the landscape such that the center of the radiating landscape shifts by $\mu_\textrm{b}/(2n_\textrm{b} - \coordination)$ in the $\ebr$-direction and $-\mu_\textrm{b}/2n_\textrm{b}$ in the $\ebb$-direction. This skew in the center of the radial ISD versus the center of the coordinate system ($\ebb,\ebr = 0,0$) is also observed in simulations\cite{Sue2015}, and indicates that, given a set of interaction energies, one may be able to predict not only the assembly's expected plateau/inherent structure, but also the solution conditions (range of $\mu_b$) within which an out-of-equilibrium assembly's magic ratio will remain the same.

\section{Ending remarks}
Simple local rules of an interaction network, along with simple energetic considerations, have allowed for the charting of regions (plateaux) of the two-component inherent structure diagram (ISD), within which identical outcomes occur. This work allows for the characterization of the classes of inherent structures that inhabit those plateaux, and re-establishes that local connectivity (degree) explains much in the inherent structure world. However, global connectivity is finally needed to establish exactly what magic number ratios are described by the inherent structures within each plateau, which I leave open as a future line of research. This work may eventually be useful in the rational design of a range of compositional polycrystalline materials formed far from equilibrium.

\begin{acknowledgments}
I thank Stephen Whitelam for discussions, and Alana Canfield Mannige and Anthony M. Frachioni for a detailed reading of the manuscript. 
This work was done at the Molecular Foundry at Lawrence Berkeley National Laboratory (LBNL), supported by the Office of Science, Office of Basic Energy Sciences, of the U.S. Department of Energy under Contract No. DE-AC02-05CH11231. 
\end{acknowledgments}


%

\end{document}